\documentclass[10pt,twocolumn]{article}
\usepackage{latex8, times, mathptmx}
\usepackage{graphicx}

\begin{document}

\title{Performance Evaluation of Multiple TCP connections in iSCSI}

\author{
Bhargava Kumar K \and Ganesh M. Narayan \and K. Gopinath \and \\
\textit{\{bhargava, nganesh, gopi\}@csa.iisc.ernet.in}\\
\textit{Computer Architecture and Systems Laboratory} \\
\textit{Dept. of Computer Science and Automation} \\
\textit{Indian Institute of Science, Bangalore.}
}
\maketitle

\begin{abstract}
Scaling data storage is a significant concern in enterprise systems and Storage
Area Networks (SANs) are deployed as a means to scale enterprise storage. SANs
based on Fibre Channel have been used extensively in the last decade while iSCSI
is fast becoming a serious contender due to its reduced costs and unified
infrastructure. This work examines the performance of iSCSI with multiple TCP
connections. Multiple TCP connections are often used to realize higher
bandwidth but there may be no fairness in how bandwidth is distributed. We
propose a mechanism to share congestion information across multiple flows in
``Fair-TCP'' for improved performance. Our results show that Fair-TCP
significantly improves the performance for I/O intensive workloads.
\end{abstract}

\section{Introduction}

\begin{figure}
  \includegraphics[height=1.5in,width=3in]{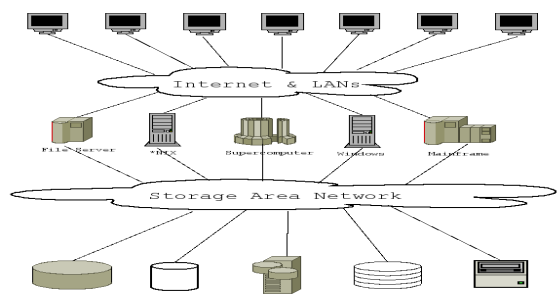}
  \caption{Elements and Ecosystem of an enterprise SAN}\label{fig:san}
\end{figure}

Future computer systems are required to scale to large volume of data that is
being generated and used, with increasing capacities and dropping prices of
magnetic disks. Traditional DAS based architectures, based on parallel SCSI
transport, scale poorly owing to their distance, connectivity and throughput
limitations and are being replaced by networked storage systems like SAN.
Figure \ref{fig:san} shows a SAN connecting multiple servers to multiple
targets.

SANs, where the storage devices are connected directly to a highspeed network, can provide high scalability and throughput guarantees; SANs allow
any-to-anywhere access across the network, using interconnect elements such
as routers, gateways, hubs and switches; they also facilitate storage sharing
between possibly heterogeneous servers to improve storage utilization and
reduce downtime.

Entities in a SAN, both storage and servers, communicate using SCSI commands.
A sender encapsulates SCSI commands over a transport protocol and sends
it to one or more receivers; receivers receive the payload, decapsulates
the commands, and execute them. Thus a SAN is defined by the transport it
uses and the encapsulation standard it follows. In this lieu, there are two
competing industry standards -- FC and iSCSI, which allow us to build SANs, each
based on differing transport and encapsulation standards.

The Fibre Channel (FC) is a serial interface, usually implemented with
fibre-optic cable. FC Standard \cite{FC} covers the physical, link, network and
transport layers of the OSI network stack and also provide a SCSI encapsulation
protocol -- FCP. FC SANs, with most FCP implementations being hardware
accelerated, provide better throughput guarantees. However, FC installations
are costlier and are cannot be deployed over long distances. Fibre Channel
requires custom network components and is not able to take advantage of the
steep technology curves and dropping costs of IP-based networks.

\begin{figure}[h]
\includegraphics[width=0.45\textwidth]{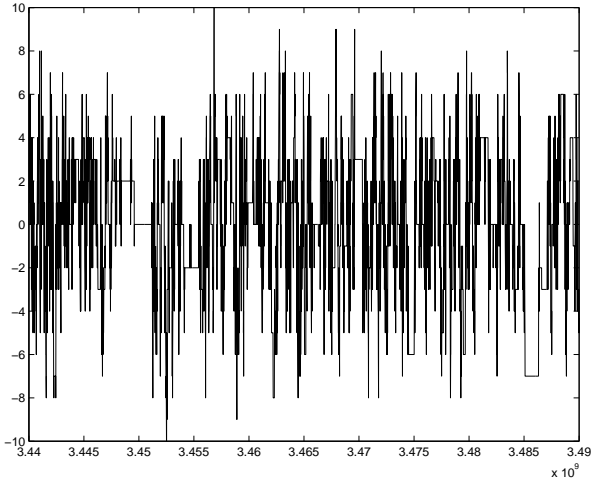}
\caption{Difference in Congestion window of 2 different connections}\label{fig:diff1}
\end{figure}

Internet SCSI or iSCSI \cite{iSCSI} is a storage networking
standard that transports SCSI commands over TCP/IP, essentially
tunneling storage protocols on top of TCP, hence IP, to leverage the installed
equipment base. This allows iSCSI to be used over any TCP/IP
network infrastructure with the remote device being seen by the
operating system as a locally available block-level device. Unlike
Fibre Channel, iSCSI can run on any network infrastructure that
supports TCP/IP. A network that uses iSCSI needs only one network
for both storage and data traffic whereas Fibre Channel requires
separate infrastructure for storage and data traffic. However,
a response to a block-level request in iSCSI may encounter a greater
delay compared to Fibre Channel, depending on the network
conditions and the location of the target.

Current efforts to improve end-to-end performance for TCP are taking
advantage of the empirically discovered mechanism of striping data
transfers across a set of parallel TCP connections between a sender
and receiver to substantially increase TCP throughput. However, when multiple
connections are used between the same source-target pairs, the connections
themselves interact/compete with each other in non trivial ways. In order to
achieve optimal throughput it is imperative that we understand these
interactions and treat the connections accordingly; failing which could lead to
increased congestion and reduced throughput.

\begin{figure}[h]
\includegraphics[width=0.45\textwidth]{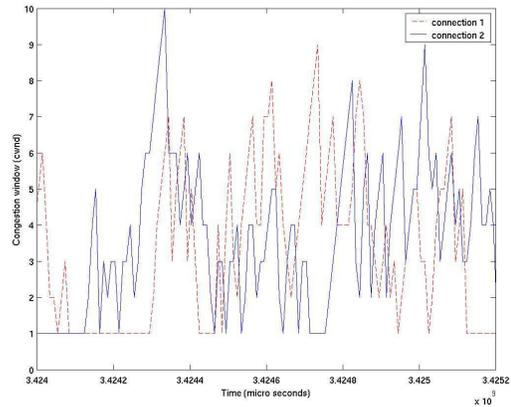}
\caption{Congestion window of 2 different connections}\label{fig:problem1}
\end{figure}

In our work, we study the effects of using multiple TCP connections on
iSCSI. It was shown in \cite{girish} that the aggregate iSCSI
throughput increases with the increase in number of TCP connections
in an emulated wide area network. We find that the multiple TCP
connections used by iSCSI compete with each other and result in
lesser throughput for iSCSI than they are capable of. We propose a
solution named Fair-TCP based on TCP control block interdependence
\cite{touch} for managing TCP connections. We compare the
performance of our variant with the standard TCP Reno\cite{reno}
with SACK\cite{sack}, using various workloads and varying delays in
an emulated wide area network. We find that for I/O intensive
workloads such as sequential write to a large file, Postmark and
Bonnie, Fair-TCP provides significant performance improvements over
standard TCP-Reno with SACK. 

Section 2 describes the behaviour of multiple TCP connections and its
effects on iSCSI. The proposed solution is also outlined there. 
Section 3 details the experimental setup, tools and
benchmarks used in our experiments. Section 4 presents our results
with a discussion. Section 5 reviews related work. Section 6 concludes
the paper.

\section{iSCSI and TCP}
SCSI standard assumes that the underlying transport is reliable and supports
FIFO ordering of commands. TCP has mechanisms to acknowledge the
received TCP packets and to resend/request packets that are not acknowledged
within a certain time period, effectively guaranteeing reliable and in-order
delivery of packets. Choosing TCP as a transport is thus a natural choice. If
iSCSI were defined on top of a protocol that is not reliable and in-order then
iSCSI would have had to provide these services by itself.

iSCSI initiators are usually connected to iSCSI targets using multiple TCP
connections. The reason is two fold: due to TCP window size restrictions and
round trip times over long distances, it might not be possible for a single TCP
connection to utilize the full bandwidth capacity of the underlying link;
secondly, there may also be several physical interconnects connecting the
initiator and target, and it would be most desirable to aggregate and
simultaneously utilize all such existing physical interconnects. As TCP does
not support such aggregation, an iSCSI session is therefore defined to be a
collection of one or more TCP connections between the initiator and the target.

\begin{figure}[t]
  \centering
  \includegraphics[width=0.45\textwidth]{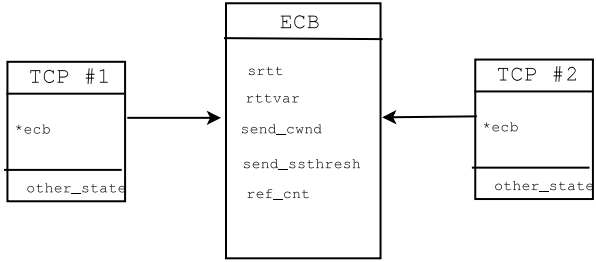}
  \caption{Fair-TCP Design}\label{fig:fair}
\end{figure}

\subsection{Behaviour of multiple TCP connections}
Many applications use multiple TCP connections between client and server for
increased throughput. However these TCP connections are treated independently:
most TCP implementations keep state on a per-connection basis in a structure
called TCP control block (TCB) or an equivalent construct and each of these TCP
connections are handled independently. Several researchers
(\cite{busyinternetserver}, \cite{integrated}, \cite{touch}) have shown that
concurrent connections such as these compete with each other for link
bandwidth, often resulting in unfair and arbitrary sharing of
bandwidth. Concurrent connections do not share indications of
congestion along the shared path between the sender and receiver.
Therefore each connection independently pushes the network to a
point where packet losses are bound to happen. Once the network is congested,
all the competing connections reduce their transmission windows drastically,
thus limiting the effective bandwidth available to the application, This
results in under utilization of the shared link, and hence less aggregate
throughput than the application is capable of achieving. Also, it often happens
that some of the connections stall due to multiple losses, while others proceed
unaffected. Thus concurrent TCP connections, when left without any explicit
arbitration, provide neither bandwidth utilization nor fairness.

Some of the information in a TCB, like round-trip time (RTT), is not
application specific but is specific to a host (or subnet). If there are
multiple TCP connections between the same hosts, each will independently monitor
its transmissions to estimate the RTT between the hosts. Such a scheme is
wasteful as it needs extra processing and memory at a TCP endpoint. An
alternate scheme is to share this information between such concurrent
connections.

In order to see if iSCSI suffers from any of the above problems, we
evaluated the performance of iSCSI with multiple TCP connections. We
observed traces of congestion window of each connection, for a
sequential file write of 1GByte to see if the connections are
competing with each other. Figure \ref{fig:problem1} shows a sample
of traces of congestion window for 2 different connections in a WAN
environment with delay of 4ms for a period of 1.2 seconds. The
congestion window was collected approximately every 10ms. Figure
\ref{fig:diff1} shows a sample of the difference in congestion
window of the 2 connections for a period of 50 seconds.

From the traces, we can see the two connections compete for
bandwidth resulting in one connection using the network more than
the other. The observed mean and standard deviation for congestion
window of the two connections are 3.38/2.14 and 3.38/2.13. The
observed mean and standard deviation for the difference in window
sizes shown in figure \ref{fig:diff1} is 0 and 3.06. The mean 0 in the
window difference indicates that over long periods each connection
gets the same amount of network bandwidth. The larger deviation in
window difference compared to the deviations in each connection's
window, indicates that when one connection has a large window the
other connection has a smaller window. This is a very undesirable
behaviour from the TCP connections which results in reduced
throughput. Similar patterns were observed in all traces. For the above
traces the mean turnaround time was 453 ms with a standard deviation
of 325ms which we try to reduce. Thus we understand that the multiple
connections between iSCSI do compete and share the bandwidth
disproportionately and underutilized the resources. In our work, we share the
congestion information among the different TCP connections to reduce the
command turnaround times and increase the throughput of iSCSI.

\begin{table}
\begin{center}
\begin{tabular}{|rcl|}
  \hline
  \multicolumn{3}{|c|}{\bf{Ensemble Allocation}} \\ \hline
  conn\_srtt & = & ecb\_srtt \\ \hline
  conn\_rttvar & = & ecb\_rttvar \\ \hline
  conn\_snd\_cwnd & = & ecb\_snd\_cwnd/ref\_cnt \\ \hline
  conn\_snd\_ssthresh & = & ecb\_snd\_ssthresh/ref\_cnt \\ \hline

\end{tabular}
\caption{Ensemble Allocation}\label{table:ensembleallocation}
\end{center}
\end{table}


\subsection{Fair-TCP}
Several researchers have worked on sharing the congestion
information among multiple TCP
connections (\cite{touch}, \cite{ensemble}, \cite{cm}). Touch \cite{touch}
proposed sharing TCP state among similar connections to improve the
behaviour of a connection bundle. A bundle of TCP connections
sharing TCB information is called an \emph{ensemble}. We have
implemented a congestion information sharing mechanism, Fair-TCP
based on Touch\cite{touch}. The TCBs of individual connections are
stripped of RTT and congestion control variables. Instead, they now
simply contain a reference to the Ensemble Control Block (ECB), of
the ensemble they are part of. Fair-TCP does not support caching of
TCB states, since connections in an iSCSI session are persistent for
a very long time, and are not reestablished frequently. Figure
\ref{fig:fair} outlines the design of Fair-TCP.

Fair-TCP aggregates congestion window and slow start threshold
values in the ECB per ensemble. Ensemble allocates fair share of
available window to each connection. Fair-TCP shares the round trip
time information among connections of the ensemble. Fair-TCP
maintains a reference count of the number of connections in the
ensemble. Table \ref{table:ensembleallocation} outlines the
allocation of congestion information to connections of the ensemble.

For each window update received from a connection, the aggregate
window is adjusted appropriately. The most recent value of {\it srtt}
(smoothed round trip time) and {\it rttvar} (round trip time variance)
reported by a connection is maintained in the ensemble. Whenever a new
connection is established, it is added to the corresponding ensemble
without any changes to the ensemble. If there is no ensemble
corresponding to that connection, a new ensemble will be created and
initialized with the values from that connection.  Fair-TCP has been
implemented on both the target and the initiator.

%

\section{Experimental Setup}

\subsection{Tools and Benchmarks}

The \emph{UNH-iSCSI }\cite{unh} protocol implementation of initiator
and target is used for all our experiments. It is designed and
maintained by UNH InterOperability lab's iSCSI Consortium. The
implementation consists of initiator and target drivers for Linux 2.4.x
and 2.6.x kernels. It supports multiple sessions between a given
initiator target pair, multiple connections per session, arbitrary
number of outstanding R2Ts, all combinations of initialR2T and
ImmediateData keys, arbitrary values of data transfer size related
iSCSI parameters and Error Recovery level 1.

The \emph{NIST Net} \cite{nistnet} network emulation tool for GNU/Linux
is used for introducing delays. NIST Net allows a GNU/Linux PC set up as
a router to emulate a wide variety of network conditions. The tool
is designed to allow controlled, reproducible experiments for
network performance sensitive/adaptive applications and control
protocols in a simple laboratory setting. By operating at the IP
level, NIST Net can emulate the critical end-to-end performance
characteristics imposed by various wide area network situations
(e.g., congestion loss) or by various underlying subnetwork
technologies. The tool allows an inexpensive PC-based router to
emulate numerous complex performance scenarios, including: tunable
packet delay distributions, congestion and background loss,
bandwidth limitation, and packet reordering/duplication.

\emph{Bonnie++} \cite{bonnie++} is a benchmark suite that is
aimed at performing a number of simple tests of hard drive and file
system performance. The benchmark tests database type access to a
single file (or a set of files), and it tests creation, reading, and
deleting of small files which can simulate the usage of programs
such as Squid, INN, or Maildir format email. The first six tests
include per-char write, block write, block rewrite, per-char read,
block read and random seeks. For each test, Bonnie reports the
number of Kilo-bytes processed per elapsed second, and the \%cpu
usage (sum of user and system). The next 6 tests involve file
create/stat/unlink to simulate some operations that are common
bottlenecks on large Squid and INN servers, and machines with tens
of thousands of mail files in /var/spool/mail.

    The \emph{Postmark} \cite{postmark} benchmark models the workload
seen by a busy web server and is sensitive to I/O latency. The
workload is meant to simulate a combination of electronic mail,
netnews and web-based commerce transactions. Postmark creates a
large number of small files that are constantly updated. Once the
pool of files has been created, a specified number of transactions
occur. Each transaction consists of a pair of smaller transactions,
i.e. Create file or Delete file and Read file or Append file. Each
transaction type and files it affects are chosen randomly. On
completion of each run a report is generated showing metrics such as
elapsed time, transaction rate, total number of files created, read
size, read throughput, write size, write throughput and so on. The
Postmark configuration used in our experiments is listed in Table
\ref{table:postmark} and rest of the parameters have been set to
default.

\begin{figure}
  \centering
  \includegraphics[width=0.45\textwidth]{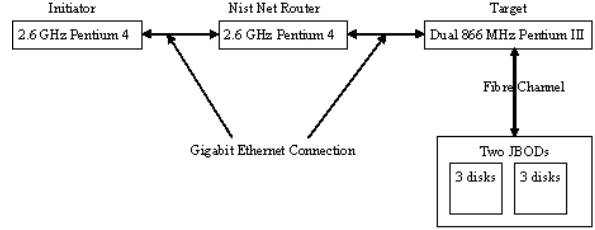}
  \caption{Experimental Testbed}\label{fig:testbed}
\end{figure}

\subsection{Experimental Testbed} Our experimental WAN emulation
testbed is illustrated in Figure \ref{fig:testbed}. Three machines
were used in our experimental setup: initiator, router and target.
All the three machines were connected to a D-Link DGS10008TL gigabit
switch using gigabit NICs.

    The initiator hosted a 2.6 GHz Intel Pentium 4 processor, 256 MBytes
of RAM and Broadcom BCM5700 gigabit Ethernet controller. It was
running the Linux 2.4.20 kernel. The system in which the target was hosted had
a dual 866 MHz Pentium III processor, 756 MBytes RAM and Fibre
Channel host bus adapter. The target was connected to two JBODs,
each housing three Seagate ST336752FC 15K RPM disks. The target was
running a Linux 2.6.5 kernel for i686. The machine designated as router
hosted a hyper-threaded 2.6 GHz Pentium 4 processor, 1 GB of RAM and
two gigabit NICs (D-Link DL2K and Intel 82547EI).

Both the initiator and the target were running UNH iSCSI implementation.
The machine designated as router between the initiator was running
NIST Net network emulation tool to simulate a WAN environment. The
WAN simulation was tuned in accordance with profiling information
presented in Vern Paxson \cite{paxson}, which found that over long
periods network connections suffered a 2.7\% packet loss in a
wide-area network. Performance measurements were calculated using
varying delays. Socket buffer sizes on both the initiator and the
target were set to 512KBytes.
\section{Results and Discussion}
In all our experiments 4 TCP connections were used in a session
between the initiator and the target. Girish\cite{girish} identifies
that beyond 4 connections the incremental increase in throughput is
very low. Standard ethernet frame size of 1500 bytes was used in all
experiments. We did not consider using jumbo frames, since in real
systems not all components in the network path support jumbo frames.
Large socket buffers are necessary to achieve peak performance for
networks with large bandwidth-delay products. Socket buffers on both
the initiator and the target were set to maximum value of 512KBytes.

\begin{figure}
  \centering
  \includegraphics[width=0.3\textwidth]{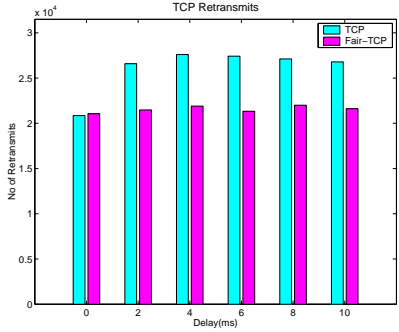}
  \caption{TCP Retransmits for block size of 1024 bytes}\label{fig:retrans}
\end{figure}

\subsection{Sequential File Writes}
\begin{figure*}
  \centering
  \includegraphics[width=0.9\textwidth]{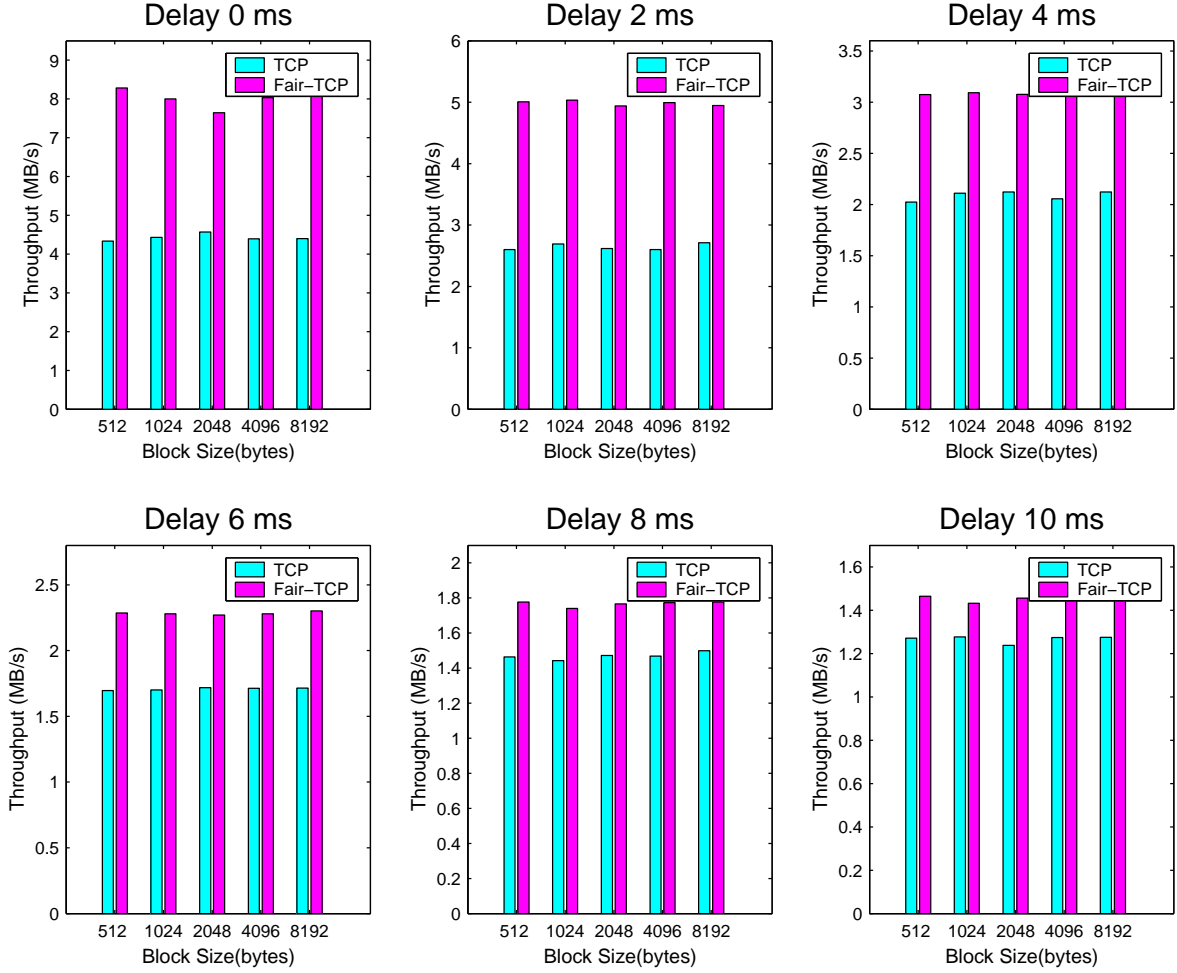}
  \caption{Throughput for Sequential File writes with varying delays and block sizes}\label{fig:write}
\end{figure*}

Figure \ref{fig:write} shows the performance of iSCSI for a
sequential file write of 1GB with different block sizes for write
system call and varying network delays. A request for \emph{fsync}
was made before closing the file to ensure that all the data were
written to the disk. Figure \ref{fig:write} shows the performance of
iSCSI with standard Reno TCP with SACK (referred to as standard TCP
or TCP) and Fair-TCP implemented on top of it. As shown in the
figure Fair-TCP performs better than normal TCP-Reno at all delays.
But as the delays increase the gap narrows, this we believe is
due to delays overwhelming the window management efficiency of
Fair-TCP.

\begin{table}[t]
\begin{center}
\begin{tabular}{|c|c|c|c|c|c|c|}
  \hline
  Delay & \multicolumn{3}{c|}{TCP} & \multicolumn{3}{c|}{Fair-TCP} \\
  \cline{2-4} \cline{5-7}
  (ms) & Mean & SD & \%SD & Mean & SD & \%SD \\ \hline
  0  & 17.0 & 5.0 & 29 & 16.4 & 3.2 & 19  \\ \hline
  2  & 13.2 & 4.3 & 33 & 16.0 & 3.4 & 21  \\ \hline
  4  & 13.6 & 4.1 & 30 & 15.7 & 3.2 & 20  \\ \hline
  6  & 14.2 & 4.1 & 29 & 15.6 & 3.2 & 20  \\ \hline
  8  & 14.5 & 4.2 & 29 & 15.6 & 3.1 & 20  \\ \hline
  10 & 14.8 & 4.1 & 27 & 15.6 & 3.1 & 20  \\ \hline
\end{tabular}
\caption{Aggregate Congestion Window}\label{table:aggrcwnd}
\end{center}
\end{table}

The block sizes used in the write system call had little effect on the
overall throughput. To find out the reason behind such behaviour, we
observed the SCSI request sizes received by iSCSI. Since the writes
were sequential the operating system was able to bundle all the file
writes into chunks of 128KBytes. The operating system aggressively
caches each write and bundles them and sends to the disk. So the
block size was not really a factor that affects the throughput for
sequential file writes.

With increasing delays throughput of iSCSI decreased rapidly. This
we believe is mainly due to the synchronous nature of iSCSI. There
can only be a limited number of pending SCSI requests that can be
held with the target. The initiator has to wait for a minimum of an
RTT seconds before sending another request, i.e. the initiator can
only send a limited number of SCSI requests during an RTT interval,
which do not generate enough traffic in the network to match the
bandwidth-delay product and get the maximum possible throughput.

To see if the increase in throughput observed in Fair-TCP is due
to better management of window or aggressive nature of Fair-TCP, we
measured the number of TCP retransmits for a block size of 1024
bytes with varying delays. The results are shown in figure
\ref{fig:retrans}. The number of retransmits for Fair-TCP are less
than that of standard TCP in almost all the cases. Fair-TCP shares
the most recent estimate of RTT, between all connections. As a
result it has fewer false retransmits than standard TCP.

Table \ref{table:write_commandtimes} shows the mean and Standard Deviation
of SCSI command turnaround times in milliseconds (ms). Mean command
turnaround times for Fair-TCP are less than that of standard TCP.
The deviation percentage is also less for Fair-TCP than standard TCP
except for the delay of 0ms, for which the deviation was more.
Further experiments are required to determine the exact reason for
such a behaviour.

Table \ref{table:aggrcwnd} shows the 1.5aggregate congestion
window of all connections for TCP and Fair-TCP for different delays
collected on the initiator (write traffic is mainly data-outs from
the initiator). Fair-TCP has a larger window and has lesser
deviation than standard TCP, which indicates Fair-TCP has more
stable window than standard TCP.

In our experiments of sequential file writes, we observe that
Fair-TCP offers better throughput and reduces the deviation in
command turnaround times. Fair-TCP is less burstier than standard
TCP and reduces the number of false retransmits. Fair-TCP ensures
that each connections an equal share of the available bandwidth.

\subsection{Sequential File Reads}

\begin{figure*}
  \centering
  \includegraphics[width=0.9\textwidth]{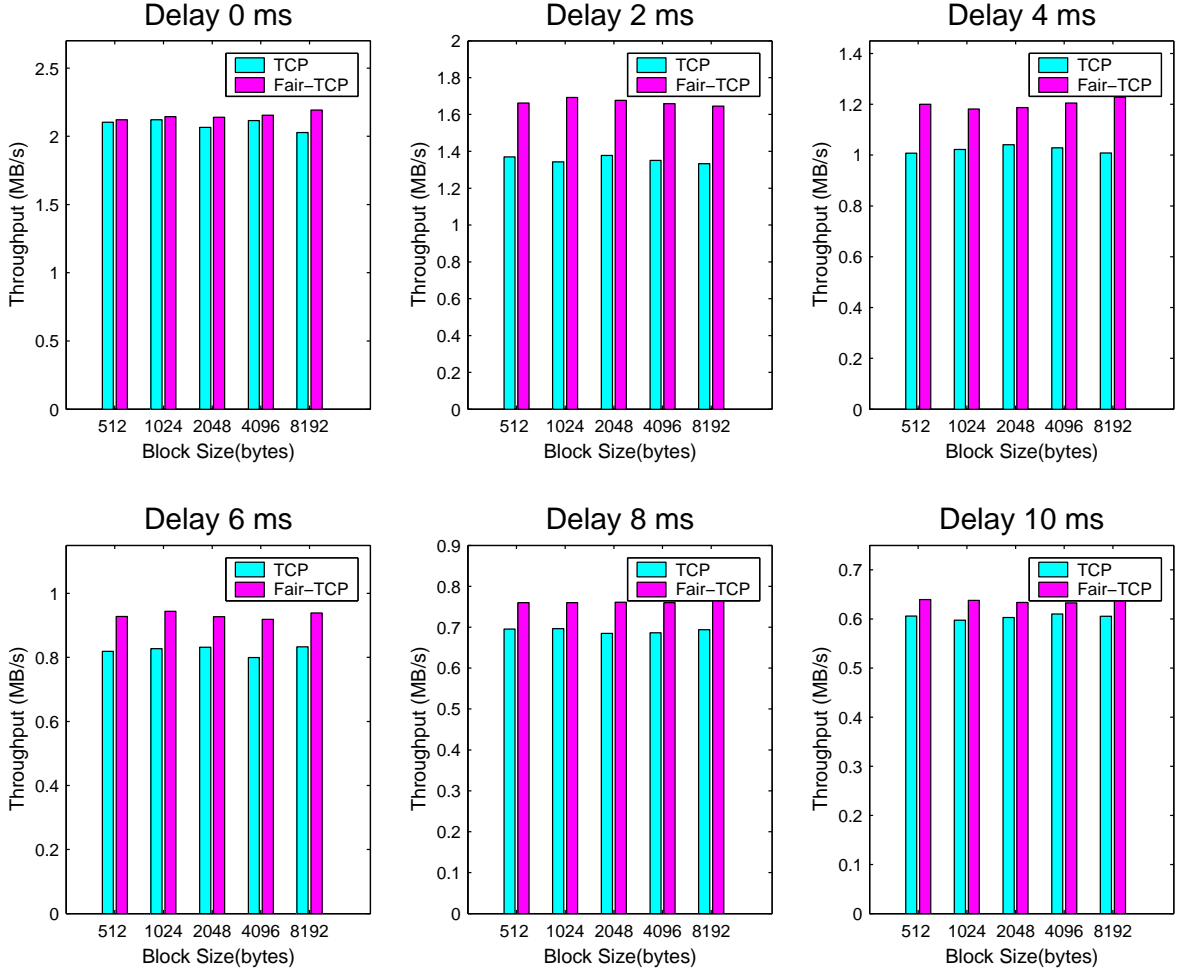}
  \caption{Throughput for Sequential file reads with varying block sizes and delays}\label{fig:read}
\end{figure*}

Figure \ref{fig:read} shows the performance of iSCSI with different
block sizes for read system call and varying network delays. The
displayed throughput are for a file read of 1GB. The throughput for
reads are less than for writes. This is mainly due to buffer cache in the
Linux kernel, which performs all the writes in the memory and flushes
them to the disk as the memory becomes full. Whereas for reads, the
operating system fetches the data from the disk whenever needed.

\begin{table}
\begin{center}
\begin{tabular}{|c|c|c|c|c|c|c|}
  \hline
  Delay & \multicolumn{3}{c|}{TCP} & \multicolumn{3}{c|}{Fair-TCP} \\
  \cline{2-4} \cline{5-7}
  (ms) & Mean & SD & \%SD & Mean & SD & \%SD \\ \hline
  0  & 208 & 225 & 108 & 102 & 131 & 127 \\ \hline
  2  & 351 & 265 & 75  & 183 & 117 & 64  \\ \hline
  4  & 450 & 288 & 64  & 310 & 146 & 47  \\ \hline
  6  & 548 & 332 & 60  & 414 & 182 & 43  \\ \hline
  8  & 642 & 376 & 60  & 414 & 182 & 39  \\ \hline
  10 & 728 & 378 & 51  & 636 & 225 & 35  \\ \hline
\end{tabular}
\caption{SCSI Command turnaround times for Writes}
\label{table:write_commandtimes}
\end{center}
\end{table}

As shown in the figure \ref{fig:read}, Fair-TCP performs better than
normal TCP-Reno at all delays. However the increase in throughput is
not significant. This is due to only one pending read request at the
SCSI layer.

\begin{figure}
  \centering
  \includegraphics[width=0.3\textwidth]{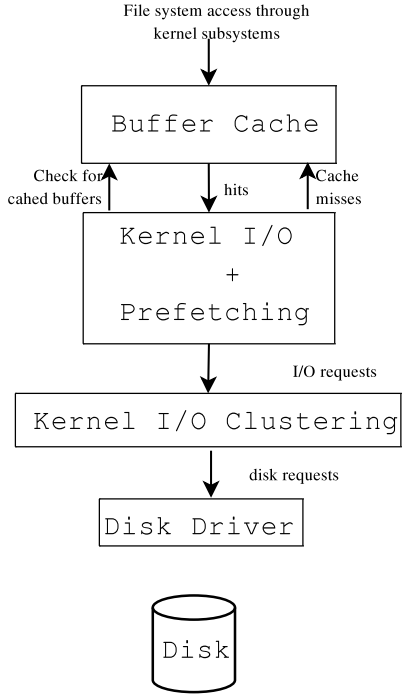}
  \caption{Kernel components involved in a read operation}\label{fig:prefetching}
\end{figure}

The block sizes used in read system call had little effect on the
overall throughput. From the traces, we observed that the read request
sent by the operating system are for 128KBytes. The Linux kernel
prefetches disk blocks starting with 1 prefetch and increases the number of
prefetches upon success. The maximum number of prefetches that can
be done is 32, which is equivalent to 128KBytes. Since the file is 
sequential, requests get clustered into a single disk read of size 
128KBytes. Figure \ref{fig:prefetching} shows the various kernel 
components involved in a read operation\cite{prefetching}.

Table \ref{table:read_commandtimes} shows the mean and Standard Deviation
of SCSI command turnaround times for requests generated in
milliseconds (ms). Mean command turnaround times for Fair-TCP are
less than that of standard TCP. The deviation percentages are also
smaller for Fair-TCP.

\begin{table}[t]
\begin{center}
\begin{tabular}{|c|c|c|c|c|c|c|}
  \hline
  Delay & \multicolumn{3}{c|}{TCP} & \multicolumn{3}{c|}{Fair-TCP} \\
  \cline{2-4} \cline{5-7}
  (ms) & Mean & SD & \%SD & Mean & SD & \%SD \\ \hline
  0  & 52  & 105 & 201 & 56  & 104 & 186 \\ \hline
  2  & 87  & 140 & 161 & 71  & 82  & 116 \\ \hline
  4  & 116 & 127 & 110 & 103 & 95  & 92  \\ \hline
  6  & 147 & 147 & 100 & 133 & 99  & 75  \\ \hline
  8  & 172 & 150 & 87  & 160 & 104 & 65  \\ \hline
  10 & 196 & 139 & 71  & 187 & 103 & 55  \\ \hline
\end{tabular}
\caption{SCSI Command turnaround times for Reads}
\label{table:read_commandtimes}
\end{center}
\end{table}

\begin{table}
\begin{center}
\begin{tabular}{|c|c|c|c|c|}
  \hline
        & \multicolumn{2}{c|}{TCP} & \multicolumn{2}{c|}{Fair-TCP} \\
  \cline{2-3} \cline{4-5}
  Delay & Rewrite & Seeks & Rewrite & Seeks \\
  (ms)  & (KB/s)  &  /sec & (KB/s)  & /sec  \\ \hline
  0  & 1383 & 65.8 & 1633 & 93.9 \\ \hline
  2  & 669  & 65.4 & 917  & 82.7 \\ \hline
  4  & 484  & 58.5 & 591  & 71.1 \\ \hline
  6  & 389  & 54.5 & 433  & 63.7 \\ \hline
  8  & 322  & 52.5 & 341  & 57.0 \\ \hline
  10 & 279  & 42.0 & 291  & 47.1 \\ \hline
\end{tabular}
\caption{Bonnie++ single process}\label{table:bonnie_1}
\end{center}
\end{table}

\subsection{Bonnie++}

Table \ref{table:bonnie_1} shows the performance of Fair-TCP and
Standard TCP for Rewrites and Seeks with Bonnie++\cite{bonnie++}. Bonnie++
was run in fast mode in all our experiments. File size of 1GB and
block size of 1024 bytes were used. Results for block writes and
block reads are omitted, since single process Bonnie++ writes and
reads are similar to the sequential file writes and reads which were
discussed before. Create/stat/unlink tests of bonnie++ were not
used, since they were similar to workload generated by Postmark.
Fair-TCP improves the performance of rewrites by about 5-35\%.
Rewrites are similar to reads, except that they are dirtied and
written back. The lower throughput seen for rewrites is mainly due
to blocking during read requests. Fair-TCP improves the performance
of seeks by 10-40\%. Due to parallel seeks (3 seeks by default),
more data is queued at the SCSI layer and results in better seek
rate for Fair-TCP.

\begin{figure}
  \includegraphics[width=0.5\textwidth]{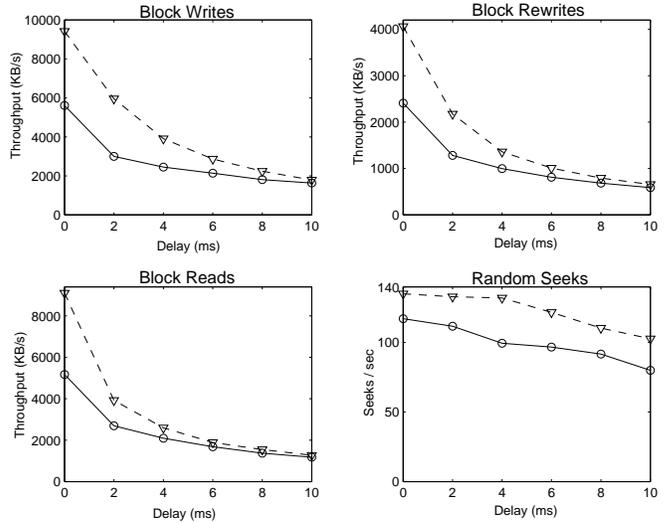}
  \caption{Bonnie++ 4 process}\label{fig:parallelbonnie}
\end{figure}

Bonnie++ allows running several instances of it in a
synchronized way using semaphores. All the process use the semaphore
will start each test at the same time. We ran 4 process of Bonnie++
using the above functionality provided by semaphores. Each process
performs the tests with a file size of 256MB and a block size of
1024 bytes. Results for block writes, block rewrites, block reads
and random seeks are shown in figure \ref{fig:parallelbonnie}. As
observed in the previous experiments, Fair-TCP performs well but the
improvement diminishes with increasing delays. However, for random
seeks, Fair-TCP performs consistently better than standard TCP and
improves the seek rate by 20-30\% irrespective of delays.

\begin{table}[h]
\begin{center}
\begin{tabular}{|c|c|}
  \hline
  Parameter & Value \\ \hline
  Number of Simultaneous Files & 20000 \\ \hline
  Lower Bound on file size & 500 Bytes \\ \hline
  Upper Bound on file size & 100 KBytes \\ \hline
  Number of Transactions & 50000 \\ \hline

\end{tabular}
\caption{Postmark Parameters}\label{table:postmark}
\end{center}
\end{table}

\subsection{Postmark}

\begin{figure}[t]
  \includegraphics[width=0.4\textwidth]{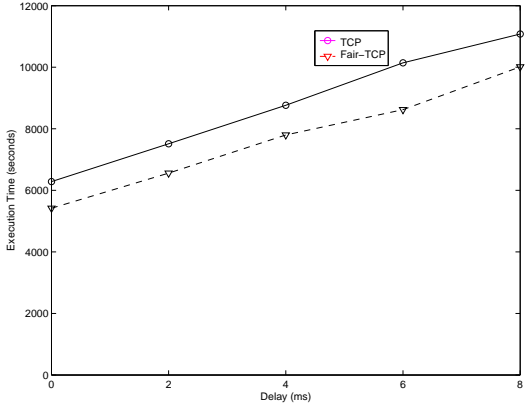}
  \caption{Postmark execution times}\label{fig:postmarktimes}
\end{figure}

Postmark\cite{postmark} was run with 20000 initial files and 50000
transactions. Figure \ref{fig:postmarktimes} shows the times taken to
run the postmark for standard TCP and Fair-TCP. Around 4GB of data was
transacted during the execution of postmark. We notice that standard
TCP needs 10-18\% more time than Fair-TCP to run Postmark.
Considering that Postmark is single threaded and reads are
synchronous, the performance improvement observed is mainly due to
asynchronous file writes and metadata writes from the buffer cache.

\begin{figure}[t]
  \includegraphics[width=0.4\textwidth]{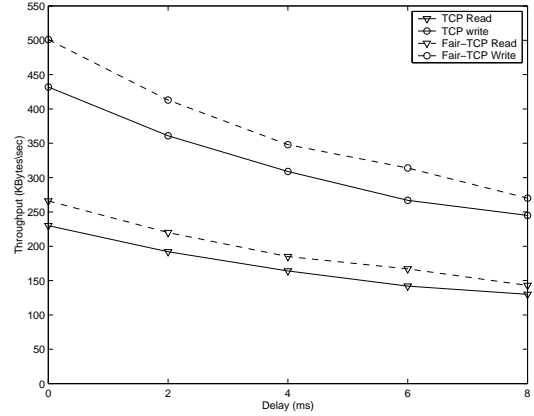}
  \caption{Postmark Read Write throughput}\label{fig:postmarkthr}
\end{figure}

Figure \ref{fig:postmarkthr} shows the read and write
throughput for standard TCP and Fair-TCP. The read and write
throughput improves by about 10-18\% for Fair-TCP. Due to filesystem
caching effects and asynchronous nature of writes, throughput for
writes in all cases is better than read throughput, and decreases
with increasing delays.

\begin{figure}[t]
  \centering
  \includegraphics[width=0.5\textwidth]{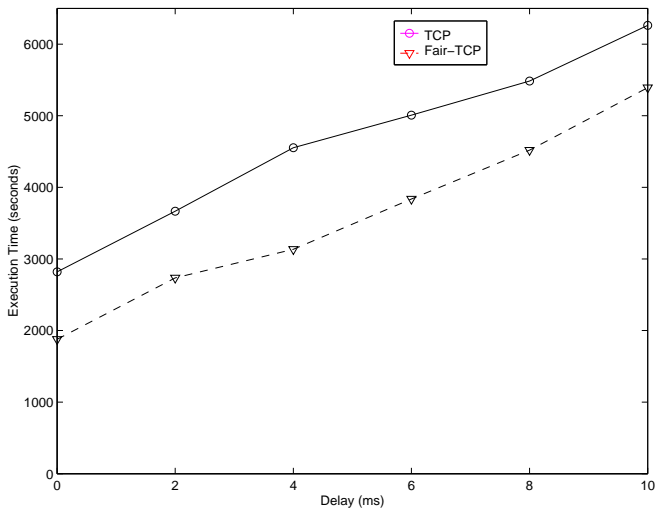}
  \caption{Execution time for 10 Postmark processes}\label{fig:parallelpostmarktimes}
\end{figure}

Postmark is completely single-threaded. In a normal web server,
there would generally be more than one thread running at a time. To
simulate the workload better we ran 10 concurrent processes of
Postmark, each with initial files of 2000 and 5000 transactions with
the rest of the parameters as in Table \ref{table:postmark}. The
results for the time take to complete all the Postmark processes are
shown in Figure \ref{fig:parallelpostmarktimes}. The times for
multiprocess Postmark are almost half that of a single process
Postmark for the same parameters. We observer that standard TCP
needs 17-50\% more time than Fair-TCP to complete Postmark
execution. The performance improvement observed going from a single Postmark
process to multiple processes is due to several requests getting
queued at the SCSI level. Fair-TCP has more data available at the
TCP level in a multiprocess environment than in a single process and this
improves the performance.

\begin{figure}
  \centering
  \includegraphics[width=0.5\textwidth]{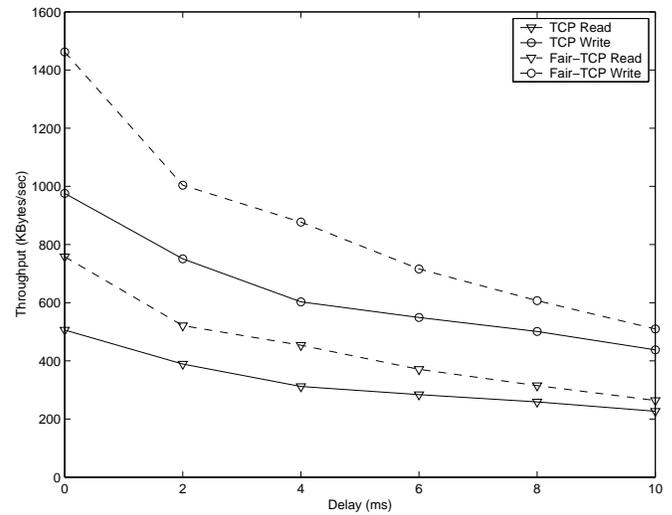}
  \caption{Aggregate Read and Write throughput for 10 Postmark processes}\label{fig:parallelpostmarkthr}
\end{figure}

Figure \ref{fig:parallelpostmarkthr} shows the read and write
throughput in a multiprocess Postmark environment for standard TCP
and Fair-TCP. The throughput shown are the aggregate throughput for
all the 10 processes. We see that Fair-TCP increases the aggregate
read and write throughput by 17-50\% over standard TCP.

\subsection{Kernel Compile}

\begin{figure}
  \includegraphics[width=0.4\textwidth]{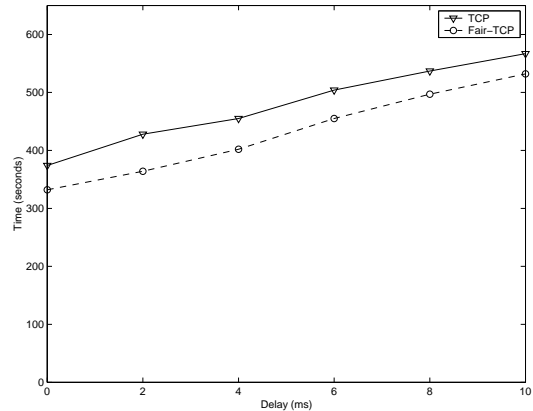}
  \caption{2.4.20 Kernel compile}\label{fig:kerneltimes}
\end{figure}

The kernel compile experiment involves untar, config and make of
2.4.20 Linux kernel. The time taken in seconds to complete the
kernel compile for various delays is shown in figure
\ref{fig:kerneltimes}. Kernel compile is CPU intensive and generates
large amounts of meta-data. Fair-TCP improves the performance of
kernel compile by 8-17\%.

\section{Related Work}

Due to increasing importance to storage scaling and reducing costs,
there has been number of efforts to build efficient implementations
of iSCSI and evaluate various aspects of iSCSI.

The work in Aiken\cite{aiken} evaluates the performance of
iSCSI in 3 different configurations, a commercial deployment of
iSCSI with Fibre Channel and iSCSI storage, a SAN environment with
software-based targets and initiators using existing hardware and in
a WAN emulated environment with varying delays. The authors in
\cite{wee} evaluate the performance of iSCSI when used in storage
outsourcing. They examine the impact of latency on application
performance and how caching can be used to hide network latencies.
The authors in \cite{farukh} use simulations to examine the impact
of various iSCSI parameters such as iSCSI PDU size, Maximum Segment
Size, Link Delay and TCP Window Size. The work in \cite{velpuri}
examines the effect of block level request size and iSCSI window
size in LAN, MAN and WAN environments. The work in \cite{girish}
examines the use of various advanced TCP stacks such as FAST TCP,
Binary Increase Congestion TCP (BIC-TCP), H-TCP, Scalable TCP and
High-Speed TCP using simulations and a emulated wan.

The authors in \cite{mirroring} study the performance of iSCSI
in the context of synchronous remote mirroring and find that iSCSI
is a viable approach to cost-effective remote mirroring. The work in
\cite{nfsvsiscsi} compares the performance of NFS and iSCSI micro
and macro benchmarks. The work in \cite{yamaguchi} examines the
impact of certain kernel SCSI subsystem values and suggest
modifications to these value for performance improvement of iSCSI.
\cite{scsi-to-ip} proposes a caching algorithm and localization of
certain unnecessary protocol overheads and observe significant
performance improvements over current iSCSI system.

\section{Conclusions}

In our work, we investigated the performance of iSCSI with multiple
TCP connections and found that iSCSI throughput suffers from competing TCP
connections. We proposed a TCB information sharing method called
Fair-TCP based on the design of \cite{touch}. We implemented
Fair-TCP for the Linux kernel and compared the performance of iSCSI with
Fair-TCP and standard TCP under different workloads. We find that
Fair-TCP improves the performance of iSCSI significantly in I/O
intensive workloads. For workloads such as single threaded read, the
SCSI data generated is quite low, hence Fair-TCP does do not as good
as in I/O intensive workloads.

\bibliographystyle{latex8}

\end{document}